\documentclass[aps,amssymb,amsmath,twocolumn,preprintnumbers,floatfix,nofootinbib,superscriptaddress]{revtex4}
\usepackage{dcolumn}
\usepackage{amsmath,amssymb}
\usepackage{epsfig,lscape,rotating}
\usepackage{graphicx}
\usepackage{color}
\usepackage{hyperref}
\usepackage{bm}
\usepackage{hyperref}
\usepackage{amsmath}
\usepackage{float}
\usepackage{array}
\usepackage{tikz}
\usetikzlibrary{calc}
\usetikzlibrary{decorations}
\usetikzlibrary{decorations.fractals}
\usetikzlibrary{decorations.pathreplacing}
 \usetikzlibrary{decorations.pathmorphing}
\usetikzlibrary{decorations.markings}
\usetikzlibrary{positioning,matrix}
\usetikzlibrary{shapes.arrows}
\usetikzlibrary{shapes.symbols}
\usetikzlibrary{shapes.misc}
\usetikzlibrary{shapes.geometric}
\usetikzlibrary{through}
\usetikzlibrary{mindmap,backgrounds}
\usetikzlibrary{fit}
\usetikzlibrary{arrows,positioning}

\include{newcommands}

\def\beq{\begin{equation}}
\def\eeq#1{\label{#1}\end{equation}}
\def\eeqn{\end{equation}}
\def\beqa{\begin{eqnarray}}
\def\eeqa#1{\label{#1}\end{eqnarray}}
\def\eeqan{\end{eqnarray}}

\def\stacksymbols #1#2#3#4{\def\theguybelow{#2}
    \def\vp{\lower#3pt}
    \def\sp{\baselineskip0pt\lineskip#4pt}
    \mathrel{\mathpalette\intermediary#1}}

\def\intermediary#1#2{\vp\vbox{\sp
     \everycr={}\tabskip0pt
     \halign{$\mathsurround0pt#1\hfil##\hfil$\crcr#2\crcr
              \theguybelow\crcr}}}

\def\to{\rightarrow}

\begin{document}

\title{ Type-III two Higgs doublet model plus a pseudoscalar confronted with $h\rightarrow\mu\tau$, muon $g-2$ and dark matter }

\author{Xuewen Liu}
\affiliation{School of Physics, Nankai University,
Tianjin 300071, China}

\author{Ligong Bian}
\affiliation{Department of Physics, Chongqing University, Chongqing 401331, China}
\affiliation{State Key Laboratory of Theoretical Physics, Institute of Theoretical Physics,Chinese Academy of Sciences,
Beijing 100190, China}
\author{Xue-Qian Li}
\affiliation{School of Physics, Nankai University,
Tianjin 300071, China}

\author{Jing Shu}
\affiliation{State Key Laboratory of Theoretical Physics, Institute of Theoretical Physics,Chinese Academy of Sciences,
Beijing 100190, China}

\begin{abstract}

In this work, we introduce an extra singlet pseudoscalar into the Type-III two Higgs doublet model
(2HDM) which is supposed to solve a series of problems in the modern particle-cosmology.
With existence of a light pseudoscalar,
the $h\rightarrow\mu\tau$ excess measured at CMS and as well as the $(g-2)_{\mu}$ anomaly
could be simultaneously explained within
certain parameter spaces that can also tolerate the data on the flavor-violating processes $\tau\rightarrow\mu\gamma$ and Higgs
decay gained at LHC.
Within the same parameter spaces, the DM relic abundance is well accounted.  Moreover, the recently observed Galactic Center gamma ray excess(GCE)
is proposed to realize through
dark matter(DM) pair annihilations, and in this work, the scenario of the annihilation being
mediated by the pseudoscalar is also addressed.

\end{abstract}

\maketitle

\section{Introduction}
\label{sec:intro}

Even though in the framework of a minimal extended Standard model (SM) with non-zero neutrino masses the leptonic flavor violation (LFV) process is
negligible for the smallness of neutrino masses which are experimentally confirmed~\cite{Vicente:2014mya}.
Therefore, a direct search for the LFV processes would provide an ideal probe for new physics beyond SM, or in other words any observational anomaly may hint us its existence.
Besides the $B$-factory, since its high energy and luminosity, LHC is definitely the machinery for the exploration. A search for LFV has been performed by
the CMS collaboration via two channels $h\rightarrow \mu\tau_e$ and  $h\rightarrow \mu\tau_h$, and
a 2.4$\sigma$ excess of the branching fraction BR$(h\rightarrow\mu\tau)=(0.84^{+0.39}_{-0.37})\%$)\cite{Khachatryan:2015kon} is reported.
If one could account such an excess as an anomaly, there should be some mechanisms which are obviously beyond SM, to be responsible for it.
The Type-III two Higgs doublet model (2HDM) is one of them, because in the model
a flavor-violating Yukawa interaction exists which may contribute to the LFV at tree level.
The model has been explored extensively~\cite{Crivellin:2013wna,Sierra:2014nqa,Crivellin:2015hha,Hektor:2015zba,Crivellin:2015mga,Omura:2015nja,Crivellin:2015lwa,Dorsner:2015mja,Altmannshofer:2015esa} to study this phenomenological observation.
Furthermore, the Yukawa interaction contributes to the muon $g-2$ via
one loop diagrams and thus would provide a possibility to explain the $(g-2)_\mu$ discrepancy~\cite{Davidson:2010xv,Omura:2015nja,Hektor:2015zba}.
Meanwhile,
the flavor-changing Yukawa interaction would induce a substantial contribution to the radiative decay $\tau\rightarrow\mu\gamma$,
thus the flavor-changing Yukawa interaction might be rigorously constrained by  the available experimental data~\cite{Davidson:2010xv,Sierra:2014nqa,Omura:2015nja}.

One of the main characteristics of the 21st century is that the cosmology has already become an accurate science and the corresponding observation must be combined
with the precise measurements and new discoveries at the facilities on the Earth to testify the standing theories. The identity of dark matter (DM) and the interaction which
determines the behavior of DM particles are the key point and searching for them is the most challenging job for both experimentalists and theorists of high energy physics and
cosmologists.

Recently, the Fermi Large Area Telescope data show
an excess of gamma-ray at energy of a few GeV coming from Galactic Center  (GCE)~\cite{Vitale:2009hr,Goodenough:2009gk,Hooper:2010mq,Hooper:2011ti,Abazajian:2012pn,Daylan:2014rsa,Zhou:2014lva,Calore:2014xka,S. Murgia}.
To explain the observation, it is suggested that annihilation of DM particles weighing 30$-$ 70 GeV
into $b\bar b$ is responsible for the GCE 
~\cite{Calore:2014nla,Agrawal:2014oha,Fan:2015sza}. Even though there exist other proposals to explain the excess,
such as a population of millisecond pulsars (MSP)~\cite{Abazajian:2010zy,Yuan:2014rca} which might be responsible for GCE,
it is not easy to explain the energy spectrum and spatial distribution of the GCE
~\cite{Hooper:2013nhl,Cholis:2014noa}. Thus in this work we discard the
astrophysical source explanation~\cite{Abazajian:2010zy,Yuan:2014rca,Hooper:2013nhl,Cholis:2014noa,Bartels:2015aea,Lee:2015fea} and focus on the DM scenario.

The dwarf galaxies are considered to be the cleanest sources for detecting gamma rays produced
by DM annihilations, thus the data on gamma ray observed at Reticulum II~\cite{Geringer-Sameth:2015lua,2015arXiv150302632T,Hooper:2015ula}
may imply existence of abundant DM at our galaxy. To be sure, we need to compute the cross section of DM annihilation in a model of particle physics.
Meanwhile other cosmological phenomena must also be concerned, namely
the DM annihilation cross section required by the new data should be of the same order as that determined by the thermal DM relic abundance.

However, the original Type-III 2HDM does not provide a natural explanation to the DM particle annihilation cross section. Thus we need to extend
the model which can also accommodate the DM annihilation.
It is note that a pseudoscalar could mediate the annihilation of dark matter(DM) pair,
meanwhile due to the small momentum transfer at t-channel,
the interaction between the DM particles coming from the outer space and
the nuclei in the detector is not affected by the existence of the new pseudoscalar, so that the DM particles may evade the direct search at not-much sensitive detectors.
This idea has been implemented in various models~\cite{Boehm:2014hva,Kozaczuk:2015bea,Cao:2015loa,Buchmueller:2015eea,Fan:2015sza,Berlin:2015wwa,Cao:2014efa,Guo:2014gra,Ghorbani:2014qpa,Huang:2014cla,
Berlin:2014tja,Agrawal:2014una,Ipek:2014gua,Berlin:2014pya,Arina:2014yna,Cline:2015qha}.
It further motivates us to consider the DM explanation to the GCE,
and introduce a Dirac fermion field serving as the DM candidate.
In this work, we introduce a pseudoscalar $a_0$ into the Type-III 2HDM.  The pseudoscalar  does not directly couple to the SM particles, but it
slightly mixes with the CP-odd Higgs which exists in the original Type-III 2HDM, thus it would effectively couple to SM via this mixing. Therefore
the pseudoscalar can mediate an effective interaction between the DM $\chi\bar\chi$ and the SM fermions $b\bar b$.

Moreover, its introduction may bring up two more advantages as following:

1. An extra pseudoscalar would open a new decay channel for the Higgs and thus
affect the  $h\to\mu\tau$ excess.

2. There could be a mixing of the newly introduced pseudoscalar with the CP-odd scalar $A_0$ in the original Type-III 2HDM, and hence induces a new contribution to
the value of $(g-2)_{\mu}$. With increase of the mixing, the contribution of the new pseudoscalar would cancel that of the Heavy Higgs, thus in this extended model the theoretical
prediction on $(g-2)_{\mu}$ can be decreased to a tolerable level.

In this work the pseudoscalar also plays a role to explain the $h\to\mu\tau$ excess and the discrepancy between theoretical prediction and data of $(g-2)_{\mu}$.

The paper is organized as follows: after this introduction, we discuss the new scenario where a light  pseudoscalar is introduced to extend the Type III 2HDM in section.~\ref{sec:model}.
In the section.~\ref{sec:phe},   we investigate the relevant topics including
the $h\rightarrow \mu\tau$ excess observed at CMS, muon $(g-2)_{\mu}$ anomaly, the galactic center gamma ray excess (GCE)
and the constraints from the
$\tau\rightarrow\mu\gamma$ process and LHC Higgs data.
The numerical  results are presented in section.~\ref{sec:results}, then the last section is devoted to our conclusion and discussion.

\section{The Model}
\label{sec:model}

In this work,  a Dirac fermion ($\chi$) of mass $m_\chi$ stands as the DM candidate and a gauge singlet pseudoscalar $a_0$ is introduced to extend  the Type-III
two Higgs doublet model, where $a_0$ mediates the coupling between
the dark matter  and the SM particles. The interaction Lagrangian reads
\begin{align}
{\cal L}_{\rm dark}&=-y_\chi a_0 \bar\chi i\gamma^5\chi\; .
\label{eq:dm-yuk-flavor}
\end{align}
The pseudoscalar $a_0$ mixes with the pseudoscalar in the original 2HDM and then couples to the SM particles through the  potential, given as\cite{Ipek:2014gua}
\begin{align}
&V=V_{\rm 2HDM}+\frac{1}{2}m_{a_0}^2a_0^2+\frac{\lambda_a}{4}a_0^4+V_{\rm portal},
\end{align}
and

\begin{align}
&V_{\rm portal}=-iBa_0H_1^\dagger H_2+{\rm h.c.}\; ,
\label{eq:Vport}
\end{align}
where $B$ is a parameter of  mass dimension, and the Higgs potential is\cite{Gunion:2002zf}
\begin{align}
V_{\rm 2HDM}&=\mu_1 H_1^\dagger H_1+\mu_2H_2^\dagger H_2+(\mu_3 H_1^\dagger H_2 + h.c.)
\nonumber
\\
&\quad+\lambda_1\left(H_1^\dagger H_1\right)^2+\lambda_2\left(H_2^\dagger H_2\right)^2
+\lambda_3\left(H_1^\dagger H_1\right)\nonumber
\\
&\quad\times\left(H_2^\dagger H_2\right)
-\lambda_4\left(H_1^\dagger H_2\right)\left(H_2^\dagger H_1\right)\nonumber\\
&\quad
+\Big[\big(\lambda_5 H_1^\dagger H_2+\lambda_6 H_1^\dagger H_1+\lambda_7 H_2^\dagger H_2\big)
\nonumber
\\
&\quad\times
\left(H_1^\dagger H_2\right)+h.c.\Big]\; .
\end{align}
We can explicitly rewrite $H_1$ and $H_2$ in the Higgs basis  as
\begin{eqnarray}
  H_1 =\left(
  \begin{array}{c}
    G^+\\
    \frac{v+\phi_1+iG^0}{\sqrt{2}}
  \end{array}
  \right),~~~
  H_2=\left(
  \begin{array}{c}
    H^+\\
    \frac{\phi_2+iA_0}{\sqrt{2}}
  \end{array}
  \right),
\end{eqnarray}
where $G^+$ and $G^0$ are the Nambu-Goldstone bosons and $H^+$ and $A_0$ are a charged Higgs boson and a CP-odd
Higgs boson, respectively. Without losing generality, let us concentrate on the CP-conserving case,
where  $a_0$ does not develop a vacuum expectation value (VEV) and all $\lambda_i$ and $\mu_3$ are set to be real.
Then the potential is minimized to
$\langle H_1\rangle=v/\sqrt2$, $\langle H_2\rangle=0$, with
$v=246~{\rm GeV}$.

In the basis of $(\phi_1,\phi_2)$,
the  mass matrix for the CP-even Higgs is ${\cal M}_h^2$ whose elements are,
\begin{align}
{\cal M}_{h11}^2&=2 \lambda_1 v^2,
\nonumber
\\
{\cal M}_{h22}^2&=m_{H^+}^2+\frac{\lambda_4 v^2}{2}+\lambda_5 v^2,
\\
{\cal M}_{h12}^2&={\cal M}_{h21}^2=\lambda_6 v^2.
\nonumber
\end{align}
Diagonalizing the matrix, one obtains the physical CP-even states
 $h$ and $H$ $(m_h\leq m_H)$ as
\begin{align}
\left(\begin{array}{c}
    \phi_1 \\
    \phi_2 \\
  \end{array}\right)
&=\left(  \begin{array}{cc}
   \cos\alpha  & \sin\alpha \\
    -\sin\alpha & ~\cos\alpha \\
  \end{array}\right)
\left(\begin{array}{c}
    H \\
    h \\
  \end{array}\right),
\\
\tan2\alpha&=\frac{2{\cal M}_{h12}^2}{{\cal M}_{h22}^2-{\cal M}_{h11}^2},
\nonumber
\end{align}
with eigen-masses being
\begin{align}
m_{h,H}^2&=\frac{1}{2}\Bigg[{\cal M}_{h11}^2+{\cal M}_{h22}^2\nonumber\\
&\mp\sqrt{\left({\cal M}_{h11}^2-{\cal M}_{h22}^2\right)^2+4\left({\cal M}_{h12}^2\right)^2}\Bigg]\; ,
\nonumber
\end{align}
and we  consider the eigenstates $h$ and $H$ as the SM-like and  heavy Higgs bosons respectively.

The CP-odd Higgs $A_0$ mixes with $a_0$ due to the potential $V_{\rm portal}$ (Eq.~\ref{eq:Vport}), and the mass matrix in the $(A_0,a_0)$ basis
is
\begin{align}
{\cal M}_A^2&=\left(  \begin{array}{cc}
    m_{A_0}^2 & Bv \\
    Bv & m_{a_0}^2 \\
  \end{array}\right),
  \end{align}
where
$m_{A_0}^2=m_{H^+}^2+\lambda_4 v^2/2-\lambda_5 v^2$.
Thus, the relation between $A_0,a_0$ and mass eigenstates $A$ and $a$ is
\begin{align}
&\quad\left(\begin{array}{c}
    A_0 \\
    a_0 \\
  \end{array}\right)
=\left(  \begin{array}{cc}
    \cos\theta & -\sin\theta \\
    \sin\theta & ~\cos\theta \\
  \end{array}\right)
\left(\begin{array}{c}
    A \\
    a \\
  \end{array}\right),
\end{align}
with the mixing angle
\begin{align}
\theta=\frac{1}{2}\tan^{-1}(\frac{2Bv}{m_{A_0}^2-m_{a_0}^2}),
\end{align}
and the masses squares are
\begin{align}
m_{a,A}^2&=\frac12\left[m_{A_0}^2+m_{a_0}^2
\pm\sqrt{\left(m_{A_0}^2-m_{a_0}^2\right)^2+4B^2v^2}\right].
\nonumber
\end{align}
The parameter $B$ in terms of $m_{a,A}$ and  $\theta$ can be expressed as
\begin{align}
B&=\frac{1}{2v}\left(m_{A}^2-m_{a}^2\right)\sin2\theta.
\end{align}
The effective coupling of the CP-even Higgs bosons to the SM $W$ is $igm_W C_{\phi WW}g^{\mu\nu}$ with $C_{hWW}=\sin\alpha,~C_{HWW}=\cos\alpha$, and
the CP-odd Higgs bosons $A(a)$ do not couple to $W$, i.e., $C_{A(a)WW}=0$. And the
$V_{\rm portal}$  is recast in terms of mass eigenstates and mixing angle as
\begin{align}
V_{\rm portal}&=-\frac{1}{4v}\left(m_{A}^2-m_{a}^2\right)
\left[s_{4\theta}~aA+s_{2\theta}^2\left(A^2-a^2\right)\right]\nonumber\\
&\quad\times
(\sin\alpha ~h+\cos\alpha ~H).
\label{eq:vport-mass}
\end{align}
The effective coupling of DM fermions to the mediator given in Eq.~(\ref{eq:dm-yuk-flavor}) is simply expressed as,
\begin{align}
{\cal L}_{\rm dark}&=-y_\chi \left(\cos\theta\,a+\sin\theta\,A\right)\bar\chi i\gamma^5\chi.
\label{eq:dm-yuk-mass}
\end{align}

The Yukawa interactions in the extended Type-III 2HDM are
\begin{align}
  {\cal L}_{\rm Yukawa}&=-\bar{Q}_L^i V_{\rm CKM}^{ij} H_1 y^i_d d_R^i -\bar{Q}_L^i V_{\rm CKM}^{ij} H_2 \rho^{ij}_d d_R^j \nonumber \\
  &\quad-\bar{Q}_L^i \tilde{H}_1 y^j_u u_R^j -\bar{Q}_L^i \tilde{H}_2
  \rho^{jk}_u u_R^k \nonumber\\
  &\quad-\bar{L}_L^i H_1 y^i_e e_R^i -\bar{L}_L^i H_2 \rho^{ij}_e e_R^j,
\end{align}
where $Q=(u_L,V_{\rm CKM} d_L)^T$, $L=(V_{\rm MNS} \nu_L, e_L)^T$ and $\tilde H_i$ stands for $i\sigma_2H_i^\ast$.
$V_{\rm CKM} (V_{\rm MNS})$ is the Cabbibo-Kobayashi-Maskawa (Maki-Nakagawa-Sakata) matrix.
The general 3-by-3 complex matrices $\rho_f^{ij}$ induce
the Higgs-mediated Flavor Changing Neutral Current (FCNC).
In the mass eigen-basis of the Higgs bosons, the Yukawa interactions are recast as
\begin{eqnarray}
  {\cal L}_{\rm Yukawa}&=&- y_{\phi i j}\bar{f}_{Li} \phi f_{Rj}
 -\bar{\nu}_{Li} (V_{\rm MNS}^\dagger \rho_e)^{ij}  H^+ e_{Rj}\nonumber \\
  &&-\bar{u}_i(V_{\rm CKM}\rho_d P_R-\rho_u^\dagger V_{\rm CKM} P_L)^{ij} H^+d_j\nonumber\\
&& +{\rm h.c.},
\end{eqnarray}
with $\phi=h,H,A,a$, $f=u,~d,~e,\nu$, and
\begin{align}\label{eq:couplings}
  y_{hij}&=\frac{m_{f}^i}{v}\sin\alpha\delta_{ij}+\frac{\rho_{f}^{ij}}{\sqrt{2}} \cos\alpha,\nonumber \\
  y_{Hij}&=\frac{m_f^i}{v} \cos\alpha\delta_{ij}-\frac{\rho_f^{ij}}{\sqrt{2}} \sin\alpha,\nonumber\\
  y_{Aij}&=\left\{
  \begin{array}{c}
    -\frac{i\rho_f^{ij}}{\sqrt{2}}\cos\theta,~(f=u),\\
    \frac{i\rho_f^{ij}}{\sqrt{2}}\cos\theta,~(f=d,~e),
  \end{array}
  \right.\nonumber\\
  y_{aij}&=\left\{
  \begin{array}{c}
    \frac{i\rho_f^{ij}}{\sqrt{2}}\sin\theta,~(f=u),\\
    -\frac{i\rho_f^{ij}}{\sqrt{2}}\sin\theta,~(f=d,~e),
  \end{array}
  \right.
\end{align}
where the couplings  $y_{Aij}$ and $y_{aij}$ exist in the new Feynman rules and are accompanied by $\gamma_5$.
For investigating $(g-2)_{\mu}$ excess, we do not need to invoke the so-called Cheng-Sher ansatz for $\rho_f^{ij}$ ~\cite{Cheng:1987rs} since
the corresponding parameter space is highly restricted~\cite{Davidson:2010xv,Sierra:2014nqa}.
The smallness of the mixing parameter $\cos\alpha$ is favored by the current LHC Higgs coupling measurements,
and we will study the issue in later part. In this scenario, the coupling of the SM-like Higgs to fermions $y_{hff}$ approaches to the SM one,
thus the flavor-violating processes mediated by the SM-like Higgs boson are almost completely suppressed.

\section{ Several relevant topics which are specifically addressed }
\label{sec:phe}

\subsection{Constraints on the parameter space of the afore model}

At first, we explore possible constraints coming from $B$ physics and Electroweak precision test, and find that the model is more advantageous over the Type-II 2HDM as it may evade those constraints in the situation of $m_{H^+}\sim m_A$ because the
$\tan\beta$ enhancement effect does not exist.

\subsubsection{Constraints from B Physics}
\label{sec:bphys}

A light  $a$ can mediate the initial and final states of decay
$B_s\to\mu^+\mu^-$ in addition to the SM contribution, hence imposes a stringent constraint on the model. For $m_a\ll m_Z$, the correction
due to an $a$ exchange at the s-channel was calculated ~\cite{Skiba:1992mg} and the results are
\begin{eqnarray}\label{Bsmumu}
&&\text{BR}\left(B_s\to\mu^+\mu^-\right)\approx \text{BR}\left(B_s\to\mu^+\mu^-\right)_{\rm SM}
\\
&&\quad\quad\quad\quad\times\left|1+\frac{v^2 m_{B_s} s_\theta^2\rho^{bb}\rho^{\mu\mu}}{4 m_\mu(m_{B_s}^2-m_a^2)}\frac{f\left(x_t,y_t,r\right)}{Y\left(x_t\right)}\right|^2,
\nonumber
\end{eqnarray}
for $\rho^{tb}_u=0$,
and
\begin{eqnarray}
&&\text{BR}\left(B_s\to\mu^+\mu^-\right)\approx \text{BR}\left(B_s\to\mu^+\mu^-\right)_{\rm SM}
\\
&&\quad\quad\quad\quad\times\left|1+\frac{v^2 m_{B_s} s_\theta^2\rho^{bb}\rho^{\mu\mu}}{ m_\mu(m_{B_s}^2-m_a^2)}\frac{f\left(x_t,y_t,r\right)}{Y\left(x_t\right)}\right|^2,
\nonumber
\end{eqnarray}
for $\rho^{tb}_u=\rho^{bb}s_\theta$,
with $x_t=m_t^2/m_W^2$, $y_t=m_t^2/m_{H^\pm}^2$, $r=m_{H^\pm}^2/m_W^2$,
and the $f$ and $Y$ functions can be found in Eq.~\ref{bsf} of the Appendix.

The average of the LHCb and CMS measurements on this mode is
BR$\left(B_s\to\mu^+\mu^-\right)=\left(2.9\pm0.7\right)\times10^{-9}$~\cite{Aaij:2013aka,Chatrchyan:2013bka,
CMSandLHCbCollaborations:2013pla}.
This could be compared against the SM prediction, which is taken to be
$\left(3.65\pm0.23\right)\times10^{-9}$~\cite{Bobeth:2013uxa,Buras:2013uqa}.
We note that
the $\tan\beta$ enhancement effect in the calculation of $B_s\to\mu^+\mu^-$ with the Type-II 2HDM~\cite{Logan:2000iv,Skiba:1992mg,Ipek:2014gua}
does not appear in our model and thus the constraint from $B_s$ leptonic decay is relaxed.

\subsubsection{The $T$ parameter}
\label{tpar}

In the Type-III 2HDM, the constraint on the $T$ parameter imposed by the electroweak precision test may be the most stringent.
Following the method proposed by the authors of Ref.\cite{Grimus:2007if,Davidson:2010xv}, the $T$ parameter in our model is obtained as,
\begin{align}
T=&\frac{1}{16\pi s_W^2 m_W^2}\Big[F(m_A^2,m_{H^+}^2)\cos^2\theta+\sin\alpha^2\nonumber\\
\times&\big(F(m_{H^+}^2,m_{H}^2)-\cos^2\theta F(m_A^2,m_{H}^2)\big)\Big]\;,
\label{Tpar}
\end{align}
with
\begin{align}
F(x,y)=\frac{x+y}{2}-\frac{xy}{x-y}\log\frac{x}{y}\;.
\end{align}

The $a/A$ and $h/H$ mixing is highly constrained by the current LHC data to be around $\sin\alpha\sim 1$ and $\cos\theta\sim1$, along with the parameter choice $m_{H^+}\sim m_A$, the $T$ parameter is suppressed, as seen
in Eq.~\ref{Tpar}, therefore it does not actually affect applications of this model as indicated in~\cite{Baak:2014ora}.

\subsection{Relevant processes under investigation}

In the following parts, we investigate several relevant processes in this extended Type-III 2HDM model. All the puzzles about the
$h\rightarrow \mu \tau$ excess, muon $g-2$ discrepancy, constraints coming from $\tau\rightarrow\mu\gamma$, dark matter relic abundance and GCE
explanations which were not solved in previous Type-III 2HDM will be addressed.

\subsubsection{$h\rightarrow \mu \tau$ excess}
\label{sec:hmt}

Existence of the flavor-violating Yukawa coupling in the extended Type-III 2HDM may possibly explain the
$h\rightarrow \mu \tau$ excess observed by the CMS collaboration.
Now let us compute the
branching ratio of $h \to \mu\tau$ in terms of our model, the result is shown as
\begin{equation}
\text{BR}(h \to \tau \mu) = \frac{m_h}{8 \pi \Gamma_h}
\left( |y_{h \tau \mu}|^2 + |y_{h \mu \tau}|^2 \right) \, ,\label{eq:hmutauBr}
\end{equation}
where $\Gamma_h$ is the total decay width of the SM-like Higgs boson.
To meet the observed excess, the flavor mixing should be of a magnitude
\begin{align}
  \bar{\rho}^{\mu\tau}&\equiv \sqrt{\frac{|\rho_e^{\mu \tau}|^2+|\rho_e^{\tau\mu}|^2}{2}}
 \simeq 0.0018 \left(\frac{\sqrt{\Gamma_h}}{\cos\alpha}  \right).
 \label{eq:hmutau}
\end{align}
For $\Gamma_h\sim 4.2$ MeV and $c_{\beta \alpha}=0.01$,  as long as $\bar{\rho}^{\mu\tau}\sim O(0.1)$
is reached, the new model is able to accommodate the $h\rightarrow \mu \tau$ excess.

\subsubsection{muon $(g-2)_{\mu}$ anomaly and $\tau\rightarrow\mu\gamma$ }
\label{sec:gmtwo}

The previous study indicated that new physics may contribute to the anomalous
magnetic moment of the muon and radiative process $\tau \to \mu \gamma$
via a chirality flipping dipole operator~\cite{Raidal:2008jk}
\begin{equation}
\frac{C^{ij}}{{\Lambda_{NP}^2}}\langle H\rangle \overline{e_i} \sigma^{\alpha \beta} P_R e_j F_{\alpha \beta} + \text{h.c.} \, ,
\label{efc}
\end{equation}
where $i,j$ denote the flavors of the external leptons, $F_{\alpha
\beta}$ is the electromagnetic strength tensor and $\sigma^{\alpha
\beta} = \frac{i}{2} \left[ \gamma^\alpha, \gamma^\beta
 \right]$.
The diagonal component
contributes to the anomalous
magnetic moment of muon, whereas the off-diagonal component corresponds to a dipole transition from $\tau$ to $\mu$.
It is noted that the coefficient $\frac{C^{ij}}{{\Lambda_{NP}^2}}$ is derived in various new physics models and has
different values which would help to determine the corresponding parameter space. In this work, we are going to derive
this coefficient in the extended Type-III 2HDM.
The flavor-violating Yukawa couplings and the newly introduced pseudoscalar induce additional contributions to
$(g-2)_\mu$ and $\tau\rightarrow\mu\gamma$ via one-loop and two loop Barr-Zee diagrams~\cite{Barr:1990vd},
as shown
in the two panels of Figure.~\ref{12loop}.
We include these extra contributions into the numerical computations.

\begin{enumerate}
\item{The muon $g$-$2$ anomaly}

In our case, the model-dependent coefficient  $C^{ij}/\Lambda_{NP}^2 $
could be expressed as~\cite{Raidal:2008jk}
\begin{eqnarray}
\frac{e \delta a_\mu}{4 m_\mu} =
\frac{{\bf Re} \{ C^{\mu \mu} \}}{\Lambda_{NP}^2}\; \cdot
\frac{v}{\sqrt{2}}.
\label{reln}
\end{eqnarray}

The contributions coming from the two CP-odd Higgs bosons give rise to
\begin{eqnarray}
\label{eq:Delamu}
 &&\delta a_\mu^{\text{1-loop}}
 \simeq
\sum_\phi  y_{\phi \tau \mu}^{2}
 \frac{ m_\mu m_\tau}{8 \pi^2m_{\phi}^2}
\left(\ln \frac{ m_{\phi}^2}{  m_\tau^2} -\frac{3}{2} \right)\;\\
 &&\delta a_\mu^{f}=-\frac{   \alpha_{\rm{em}}m_\mu }{4 \pi^3 m_f } \sum_{i,f} N_f^C Q_f^2 y_{\phi\mu\mu}y_{\phi f\bar{f}} f_\phi(r_\phi^f),\nonumber\\
 \end{eqnarray}
where $\phi = A, a$, $r_\phi^f=m_f^2/m_\phi^2$ and $y_{\phi f f'}$ is defined in
Eq.(\ref{eq:couplings}),  $f_\phi$ could be found in Eq.~\ref{opef} of the attached Appendix.
The $W$ and Goldstone loops would not contribute to the Barr-Zee diagrams for $C_{A(a) WW}=0$.

\item{ The $\tau \to \mu \gamma$ process}
\label{sec:taumugamma}

With the lepton flavor violating  Lagrangian
\begin{align}
{\cal L} = e m_l A_\mu \bar{l}_j [i \sigma^{\mu\nu}q_\nu(\mathcal{A}_L^{ij} P_L + \mathcal{A}_R^{ij} P_R) l_j] \\ \nonumber
+ h.c,
\end{align}
the coefficients $C^{ij}$ in Eq.~(\ref{efc})  could be expressed in terms of
the  form factors $\mathcal{A}_L$ and $\mathcal{A}_R$ as done in Ref.\cite{Raidal:2008jk},
\begin{equation}
  C^{\tau \mu} = \frac{e \, m_\tau \Lambda_{NP}^2 \mathcal{A}^{\tau \mu}_{R}}{\sqrt{2} v}
, \quad
  C^{\mu \tau \ast} =\frac{e \, m_\tau \Lambda_{NP}^2 \mathcal{A}^{\tau \mu}_{L}}{\sqrt{2} v}\, ,
\end{equation}
Thus, the  branching ratio of $\tau \to \mu \gamma$ is calculated as
\begin{align}
\text{BR}(\tau \to \mu \gamma) = \text{BR}(\tau \to \mu \nu \bar{\nu})\times \,\\ \nonumber
\frac{48 \pi^3 \alpha_{\rm{em}}}{G_F^2} \left( |\mathcal{A}_L|^2 + |\mathcal{A}_R|^2 \right) \, .
\end{align}

\begin{figure}[!htp]
\includegraphics[width=0.25\textwidth]{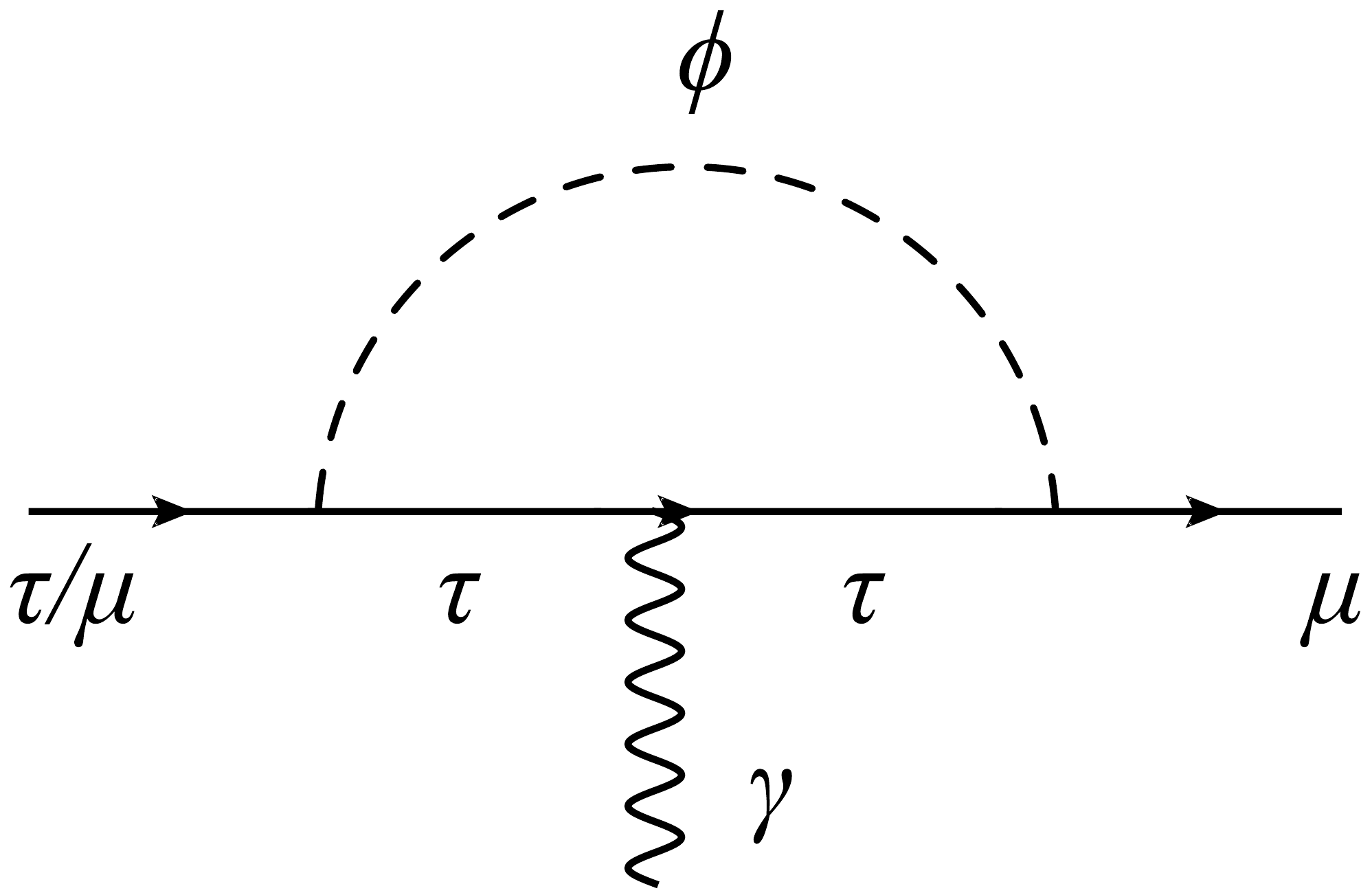}\\
\includegraphics[width=0.235\textwidth]{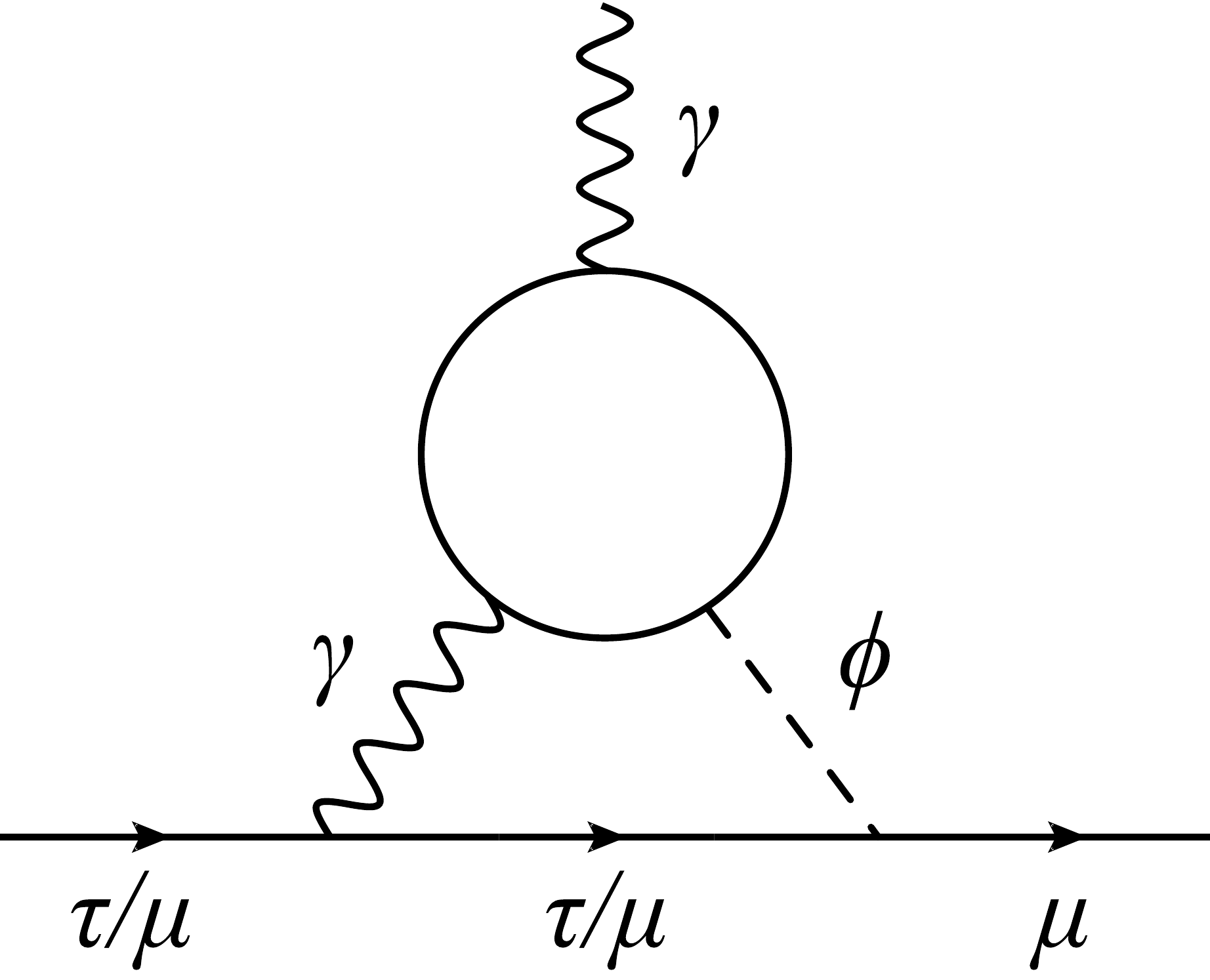}
\caption{The Feynman diagrams for $\tau\rightarrow \mu\gamma$ and $(g-2)_\mu$, where $\phi$ represents $h,H,A,a$ and
the loop at the bottom panel corresponds to $t,W$ and Goldstone loops.  }\label{12loop}
\end{figure}

Due to $|y_{h \tau \mu}| = |y_{h \mu
\tau}|$ in the model,
we have $|\mathcal{A}_L| = |\mathcal{A}_R| \equiv |\mathcal{A}|$.
The two CP-odd Higgs bosons contribute to the
form factor $\mathcal{A}$ through 1-loop and 2-loop Barr-Zee diagrams
as shown in Fig.~\ref{12loop}, and the new form factor is obtained as
\begin{align}
\label{eq:A}
\mathcal{A} &= \frac{1 }{16 \pi^2} \left( \mathcal{A}_1 + \mathcal{A}_2^{t,b} \right) \, .
\end{align}
where
\begin{eqnarray}
   \mathcal{A}_1 &=&\sqrt{2} \sum_\phi
 \frac{ y_{\phi \mu \tau}  y_{\phi \tau \tau}}{m_{\phi}^2}\left( \ln \frac{m_{\phi}^2}{m_\tau^2} - \frac{3}{2} \right) \, , \label{eq:A1}\\
\mathcal{A}_2^{t,b} &=& 2  \sum_{\phi,f} y_{\phi \mu \tau} y_{\phi ff}
\frac{N_c Q_f^2 \alpha_{\rm{em}}}{ \pi}
\frac{f_\phi(r_\phi^f) }{m_\tau m_{f}} \; .\label{eq:A2b}
\end{eqnarray}
The other  contributions from the CP-even Higgs bosons are induced by the 2-loop Barr-Zee diagrams where an
intermediate photon and a $W$-boson are involved\cite{Chang:1993kw,Davidson:2010xv}.

\end{enumerate}

\subsubsection{DM annihilation and GCE}
\label{Darkmatter}
For $m_a\ll m_A$, the Dirac DM fermions annihilate into $b\bar{b}$ primarily through exchanging $a$ at s-channel,
the annihilation cross section for the relative velocity $v_r$ is given as
\begin{equation}
\langle\sigma v_r\rangle  \simeq \frac{3 }{16 \pi } \frac{y^2_{\chi} (\rho^{ij}_f)^2 s_{2\theta}^2m_\chi^2}{(s-m_a^2)^2+m_a^2 \Gamma_a^2},
\label{DManncs0}
\end{equation}
where $\sqrt s$ is the center-of-mass energy of the annihilating DM fermions, $m_a$ and $\Gamma_a$ are the mass and decay width of the mediator boson $a$ respectively. In the non-relativistic approximation $s \sim 4m_\chi^2+ m_\chi^2 v_r^2$, thus Eq. \ref{DManncs0} can be rewritten as
\begin{equation}
\langle\sigma v_r\rangle  \simeq \frac{3 }{256 \pi m_\chi^2} \frac{y^2_{\chi} (\rho^{ij}_f)^2 s_{2\theta}^2 }{(\delta+v_r^2/4)^2+\gamma^2},
\label{DManncs}
\end{equation}
where $\gamma \equiv m_a \Gamma_a/4 m_\chi^2$ and $\delta$ are two dimensionless parameters, and the kinematic factor $\delta$ is defined as
$\delta=1-m_{a}^2/(4 m_{\chi}^2)$.

As long as $\delta$ is not too small, the DM annihilation could occur in the region far away from the resonance, then the cross section is almost independent of the velocity. In this case,
the GCE and correct thermal DM relic density could be accommodated simultaneously provided that the parameter $y^2_{\chi} (\rho^{ij}_f)^2 s_{2\theta}^2$ is adjusted to an appropriate value.
For small $\delta$, the resonance effect would enhance
the DM annihilation cross section.
For that case, adjusting the parameter $y_{\chi}^2 (\rho^{ij}_{f})^2 s_{2\theta}^2/(\delta^2+\gamma^2)$ can give a reasonable explanation to the GCE observation.
When $\delta>0$,  Eq. \ref{DManncs} indicates that the magnitude of $\langle\sigma v_r\rangle$ decreases as the temperature increases and the process
$\bar{\chi}\chi \rightarrow a \rightarrow b\bar{b}$ does not sufficiently reduce the DM abundance at the freeze-out epoch, therefore some other
annihilation processes which affect the DM relic abundance must exist in the Lee-Weinberg evolution equation.

\section{A synthesis of all the ingredients}
\label{sec:results}

In this section, we perform a numerical analysis to investigate the CMS $h\rightarrow\mu\tau$ excess, muon $(g-2)_{\mu}$ anomaly and the  $\tau\rightarrow\mu\gamma$, as well as DM relic abundance and GCE in the extended Type-III 2HDM.
The model has been implemented in the program {\tt FeynRules}\cite{Alloul:2013bka}, and the model file of the form {\tt CalcHEP}~\cite{Belyaev:2012qa} has been employed in the packages  {\tt micrOMEGAs 4.1.8} \cite{Belanger:2014vza} to calculate  the relic density and the annihilation cross section of DM.

The Higgs masses in the model are set as: $m_H=150~{\rm GeV}, m_{H^\pm}= m_A=300~{\rm GeV}$. Here the value of $m_{H^\pm}$ is
allowed by the flavor physics constraint~\cite{,Mahmoudi:2009zx}, and  $m_A=m_{H^\pm}$ is
suggested by the $T$ parameter constraint as indicated in section.~\ref{tpar}, and the value of $m_H$ employed in our computations is consistent with
the vacuum stability requirement~\cite{Ferreira:2009jb}.
The magnitude of the two CP-even Higgs mixing angle $\alpha$, is severely constrained by the recent Higgs data~\cite{Khachatryan:2014jba}, and its closeness to $\pi/2$ will be explored in later parts of this work.

The invisible and undetected decays of SM-like Higgs boson are accounted
as the decays of  beyond SM (BSM) Higgs boson \cite{Khachatryan:2014jba}, and it is denoted as
$\Gamma(h\rightarrow a a)+\Gamma(h\rightarrow \mu\tau)=\Gamma_{\text{BSM}}(h)$. Since
the contribution of BSM to the decay branching ratio is bounded bellow $0.34$ at $68\%$ CL, the decay of $h\rightarrow aa$ and $h\rightarrow \mu\tau$ in this model would be constrained. The best fit of the branching ratio of the undetected decays is $\le 0.23$ at $68\%$ CL,
and this limit allows that of BR$(h\rightarrow\mu\tau)=(0.84^{+0.39}_{-0.37})\%$, therefore it implies that the constraint on the flavor violating process is relaxed.
As $m_a < m_h/2$, the $a/A$ mixing angle $\theta$ dominates the exotic decay rate of the SM-like Higgs $h\rightarrow aa$ and changes the total decay width of the SM-like Higgs, by which the prediction value of BR$(h\rightarrow \mu\tau)$  in this model would be affected.
The magnitude of $\theta$ is required to be at order of $O(0.01-0.1)$ required by the present Higgs signal fit which is also welcome by the estimate of the DM relic abundance and GCE interpretation since this value affects the magnitude of $\langle\sigma v_r\rangle$( see Eq.\ref{DManncs}).

The parameter dependence of the coupling $\rho^{bb}_d$( denoted as $\rho_{bb}$ for simplicity) needs to be carefully analyzed for the following reasons: the coupling $\rho_{bb}$ is
responsible for one of the dominant decay modes of the SM-like Higgs boson $h\rightarrow b\bar b$,  thus any change of $\rho_{bb}$ would affect the theoretical prediction
on  BR$(h\rightarrow\mu\tau)$; meanwhile the DM annihilation $\chi\chi\to b\bar b$ is supposed to be the dominant one and its cross section is proportional to the square of $\rho_{bb}$, as given in Eq.\ref{DManncs}.  $\rho_{bb}$ is set at the same order as $\rho_{\mu\tau}$, i.e. as aforementioned in sec.\ref{sec:hmt}.
The contributions of
tau lepton and top quark dominate  BR($\tau \rightarrow \mu \gamma$), as shown in Eq.\ref{eq:A1} and Eq.\ref{eq:A2b}.
Thus the two parameters $\rho^{\tau\tau}_e$ and $\rho^{t t}_u$ (denoted as $\rho_{\tau\tau}$ and $\rho_{t t}$ for simplicity) need to be explored, here we use the parameter range of $\rho_{\tau\tau}(\rho_{t t})\sim\rho_{bb}$.  Even though we do not take the Cheng-Sher ansatz for $\rho_f^{ij}$, we still choose a negative value for $\rho_f^{ij}$ as in the Cheng-Sher ansatz~\cite{Davidson:2010xv,Sierra:2014nqa}, except for $\rho_{\tau\tau}$ and $\rho_{tt}$ while considering the current experimental constraint of $\tau\rightarrow\mu\gamma$.

To interpret GCE, the DM fermion mass and the coupling of the DM fermion to the
mediator ($a, A$) are set as $m_\chi=30~{\rm GeV}$ and $y_\chi=0.5$, and the range of $m_a$ ranges between 30 GeV-95 GeV to account for the
resonance effects in the DM annihilation process.

To obtain the parameter spaces favored by the physical picture including all the aforementioned constraints, we perform a complete numerical analysis for all possible parameter spaces: $\alpha-\rho_{\mu\tau}$, $\alpha-\theta$, $\theta-\rho_{\mu\tau}$, $\rho_{bb}-\rho_{\mu\tau}$,
$\rho_{tt}-\rho_{\tau\tau}$, and $m_a-\theta$, with relevant parameters being free within the ranges of: 30 GeV$\leq m_a\leq$ 95 GeV, $0.025<\theta<0.1$, $1.475<\alpha<1.57$, $-0.115<\rho_{\mu\tau}<0$, $-0.3<\rho_{bb}<0$, $-0.28<\rho_{tt}<0.28$, $-0.05<\rho_{\tau\tau}<0.05$ based on aforementioned arguments.
For each specific parameter space listed above, several parameters need to be fixed as shown in the following

\begin{itemize}

\item \textit{for parameters spaces of $\alpha-\theta$ ($\alpha-\rho_{\mu\tau}$)}:

 $~\rho_{\mu\mu}=-0.01, ~\rho_{\tau\tau}=0.012, ~\rho_{tt}=-0.2, ~\rho_{bb}=-0.2$, and  $m_a=50$GeV(46 GeV), $\rho_{\mu \tau}=-0.102$($\theta=0.06$) ;

\item \textit{for parameters spaces of $\rho_{tt}-\rho_{\tau\tau}$($\rho_{bb}-\rho_{\mu\tau}$)}:

 $m_a=50 $GeV, $\alpha=1.546,~\rho_{\mu\mu}=-0.01,~\theta=0.045$,
$\rho_{bb}=-0.2$ and $\rho_{\mu\tau}=-0.102$ ($\rho_{\tau\tau}=0.012$ and $\rho_{tt}=-0.2$);

\item \textit{for the $\rho_{\mu\tau}-\theta$($m_a-\theta$) parameter spaces}:

 $\alpha=1.546,~\rho_{\tau\tau}=0.015,~\rho_{tt}=-0.2,~\rho_{bb}=-0.2$, and $\rho_{\mu\mu}=-0.01(-0.02)$, $m_a=50$GeV ($ \rho_{\mu \tau}=-0.102$).

\end{itemize}

\begin{figure*}[!t]
\centering
\mbox{\hspace*{-1cm}\includegraphics[width=.39\textwidth]{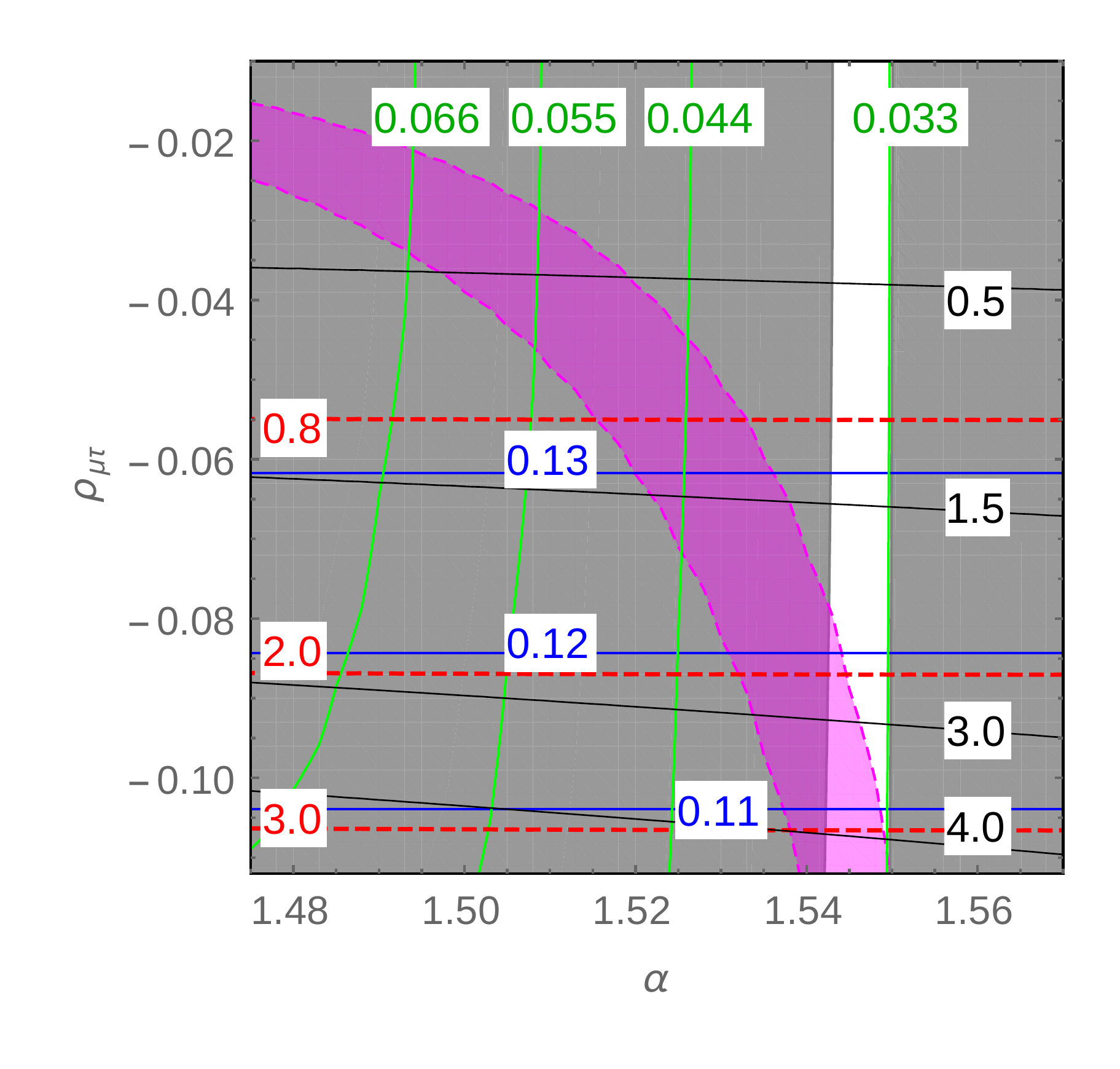}
\quad\includegraphics[width=.385\textwidth]{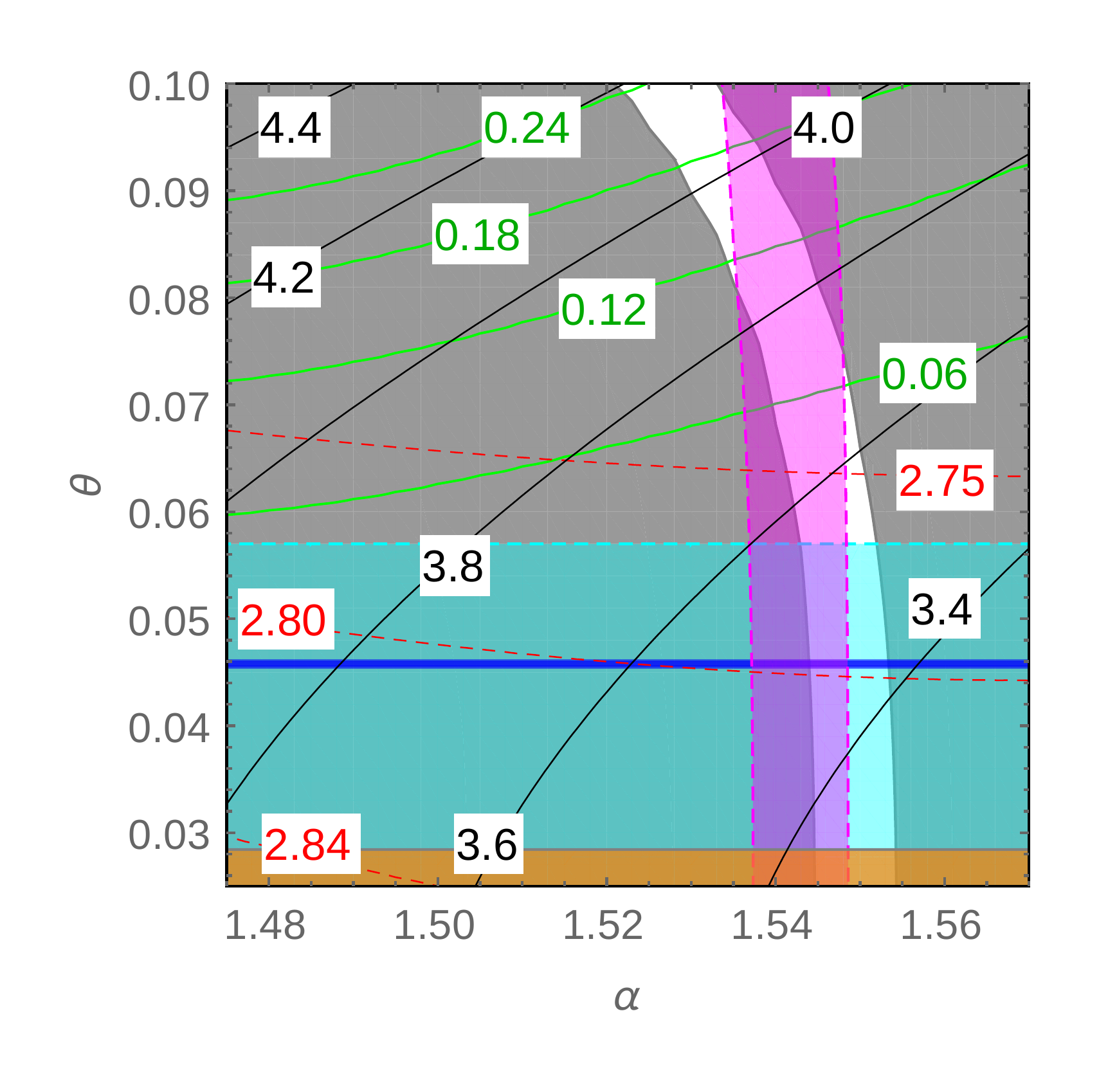}}
\mbox{\hspace*{-1cm}\quad\includegraphics[width=.39\textwidth]{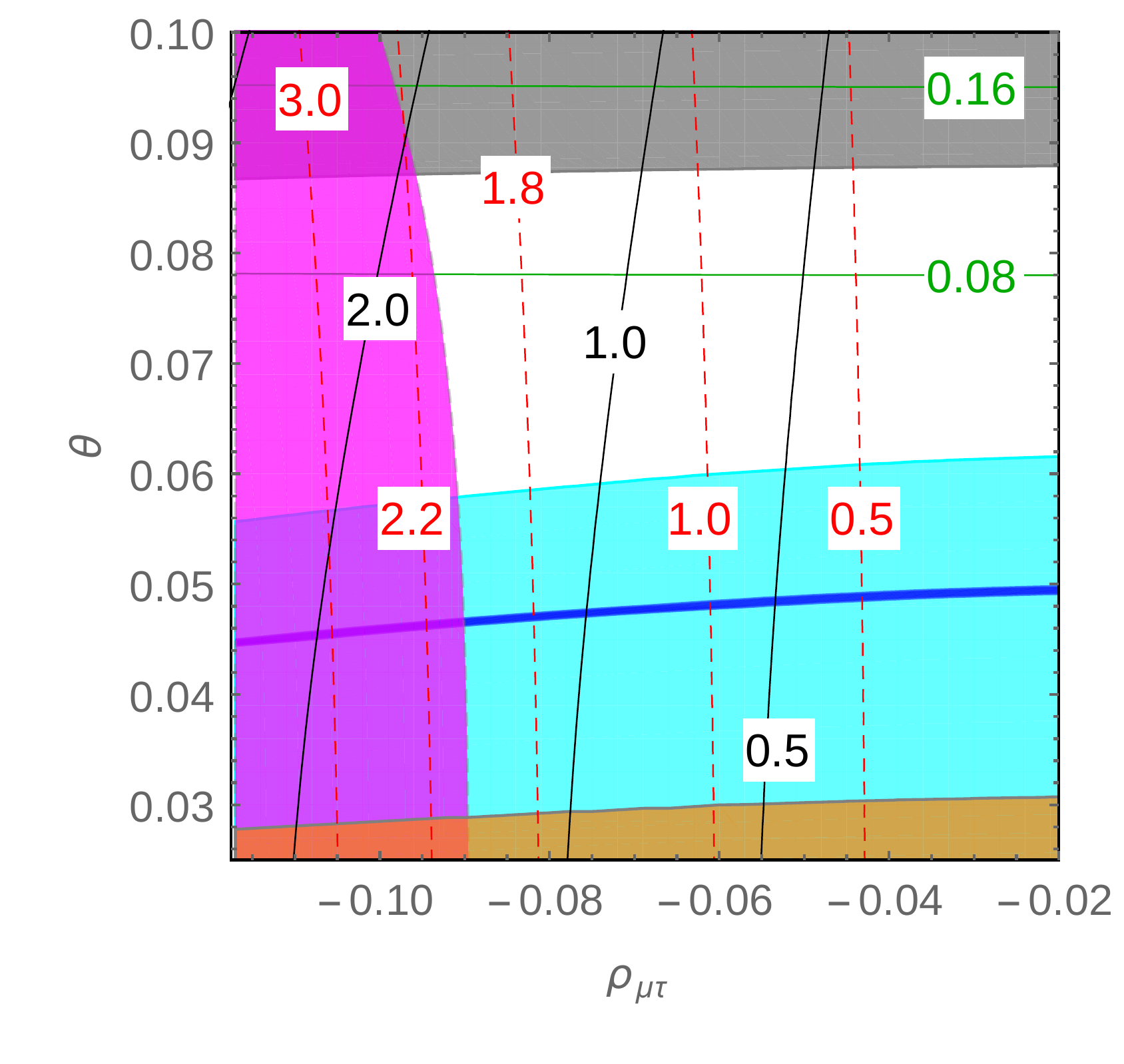}
\quad\includegraphics[width=.39\textwidth]{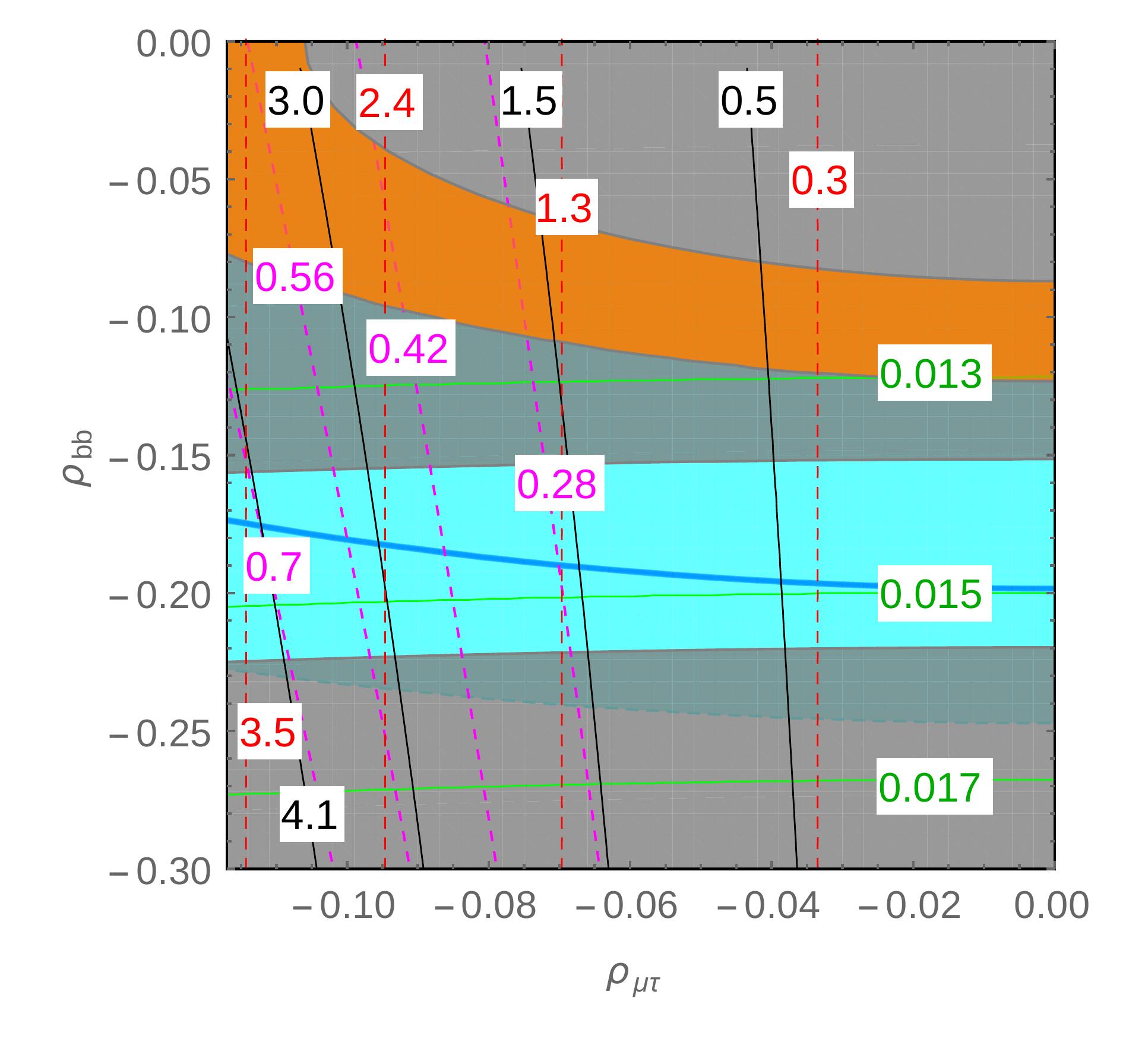}}
\mbox{\hspace*{-1cm}\quad\includegraphics[width=.39\textwidth]{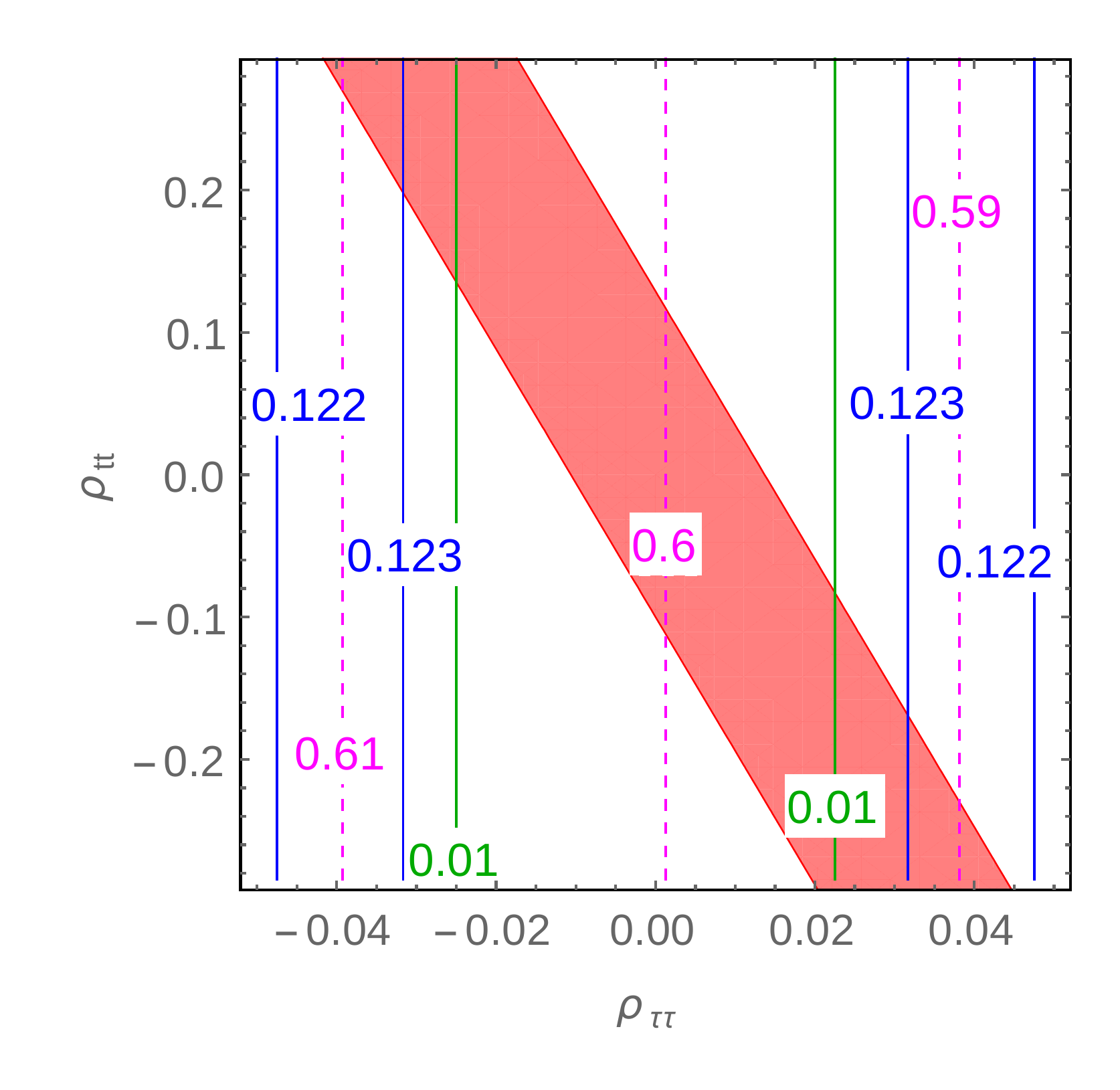}
\quad\includegraphics[width=.39\textwidth]{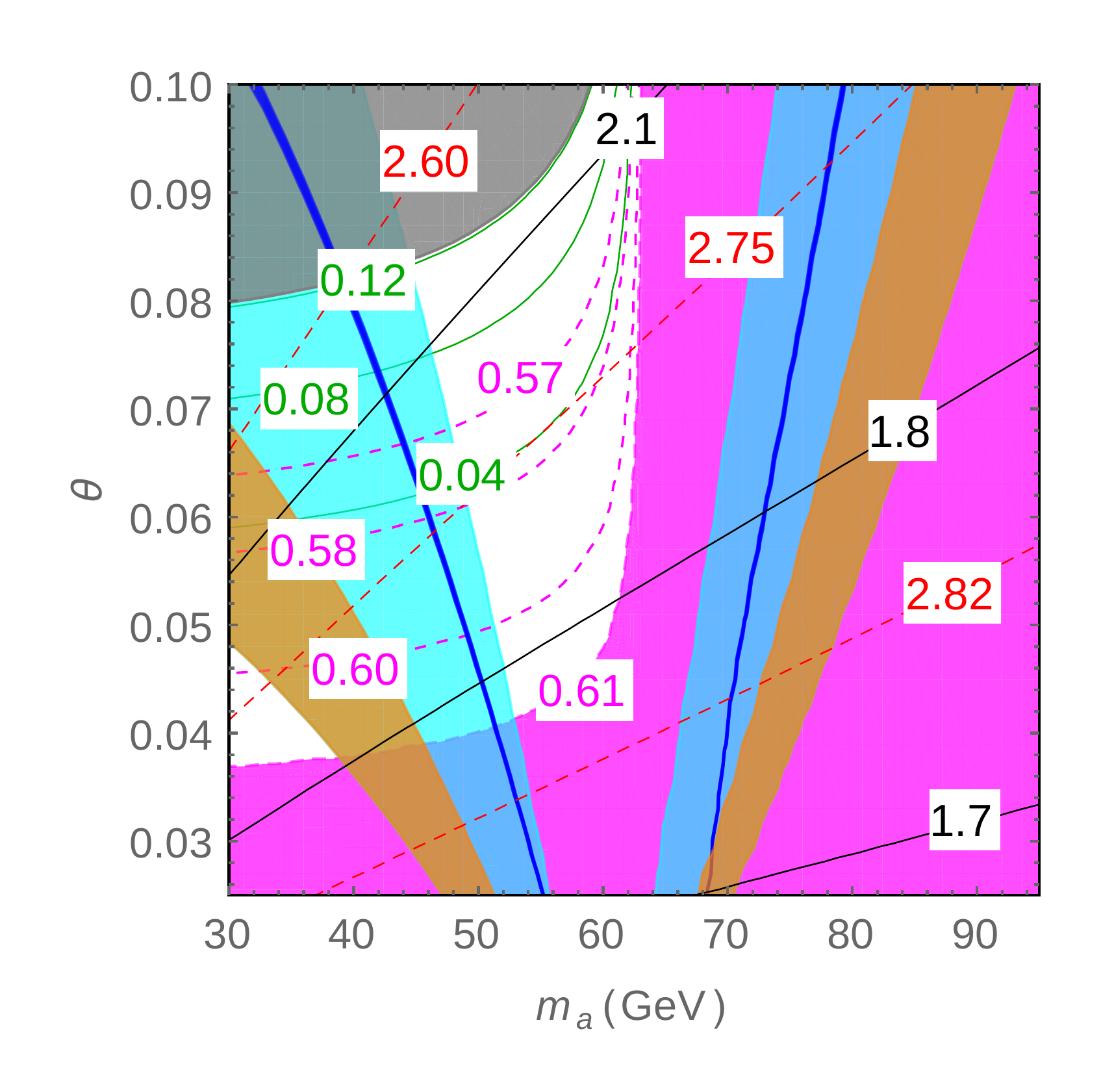}}
\caption{The corresponding results are presented as functions of
$\alpha$, $\rho_{bb}$, $\rho_{tt}$, $\rho_{\tau\tau}$,
$\rho_{\mu\tau}$, $m_a$ and $\theta$. The cyan (orange) regions give the DM annihilation cross
section of $\langle\sigma v_r\rangle
=(0.5-4)\times10^{-26}$cm$^{3}$/s($(0.5-1)\times10^{-26}$cm$^{3}$/s).
DM relic density $\Omega h^2$($=0.1197\pm0.0022$) is depicted by blue lines and contours. Magenta region/contours are for CMS $h\to\mu\tau$ excess. 
Red dashed lines are contours of $\delta a_{\mu}$/$10^{-9}$.
The black curves are contours of BR$(\tau \rightarrow \mu\gamma)$/$10^{-8}$, and the red region in the bottom-left panel is consistent with the experimental constraint $<4.4\times 10^{-8}$. The green curves are contours of BR$(h\rightarrow aa)$. The gray region is ruled out by the constraints of CMS Higgs signal strengths.}
\label{Fig:results}
\end{figure*}

Conducting a numerical analysis by means of the above parameter setup, the relevant processes are depicted  by the figure~\ref{Fig:results} (see the caption of the figure for details).

\begin{enumerate}

\item \textit{The $h\rightarrow \mu \tau$}

To understand the CMS excess BR$(h\rightarrow \mu \tau)=0.84^{+0.39}_{-0.37}\%$,
the relation between the coupling $\rho_{\mu\tau}$ and $\alpha$ given in Eq.~\ref{eq:hmutau}
should be satisfied. The region colored by magenta in the top-left plot of Fig.~\ref{Fig:results} is allowed to explain this excess. With a properly fixed $\alpha$, the both plots of the middle panel show that a sizeable coupling $\rho_{\mu\tau}\sim0.1$ is required.

The exotic decay mode $ h\to aa$ with a large
branching ratio can efficiently change the total
Higgs decay width, thus the $a/A$ mixing angle $\theta$ and the mediator mass $m_a$ affect
the prediction of BR$(h\to \mu \tau)$, as plotted in the bottom-right panel.

The whole magenta region in the bottom-right panel with the
value of BR$(h\rightarrow \mu \tau)=0.61$ implies that when $\theta$
is relatively small  or in a case of $2m_\chi<m_a$, the decay
process of $h\rightarrow aa$ cannot occur, so would not affect the
Higgs total width and becomes irrelevant to $h\rightarrow\mu\tau$
process.  Eq.\ref{eq:hmutauBr} clearly interprets the situation.

It is worth indicating that when this work is close to be finalized,
the ATLAS collaboration published new analysis on $h\rightarrow
\mu\tau_{h}$, they obtained a slightly smaller excess compared to
the CMS result, while its upper bound is consistent with the CMS
result~\cite{Aad:2015gha}.

\item \textit{The muon $g-2$ anomaly}

The contribution of two-loop Barr-Zee diagrams to $\delta a_\mu$ is
negligible because of the smallness of $\cos\alpha$. The dominant
contributions to $\delta a_\mu$ include the one-loop diagrams where
CP-even Higgs $H$ and CP-odd $A$ are mediators, especially, the two
diagrams respectively provide negative and positive contributions.

The contribution of the one-loop diagram where CP-odd $a$ is the
mediator to $\delta a_\mu$ is positive, it becomes larger for
smaller $m_a$ and/or larger $\theta$, therefore cancels out the
contribution of the diagram where $H$ is involved. The situation is
depicted at the top- and bottom-right panels of
Fig.~\ref{Fig:results} (by  the red dashed curves).

To gain the outcome result which is consistent with the present
experimental value ($\delta
a_{\mu}$/$10^{-9}=(2.61\pm0.8)$~\cite{Hagiwara:2011af}), a sizeable
$\rho_{\mu\tau}$ is required, it is also consistent with the CMS
$h\to\mu\tau$ excess, and its dependence on $\rho_{\mu\tau}$ is
demonstrated in the plots(top-left and the middle panels).

Within the allowed parameter space in $\rho_{\tau\tau}$-$\rho_{tt}$,
a value of  muon $g$-$2$: $\delta a_\mu=2.80\times10^{-9}$ is
reached.

\item \textit{Constraints from $\tau\rightarrow\mu\gamma$}

The dominant contributions come from the one-loop diagrams ($A$-loop
and $H$-loop diagrams) and the Barr-Zee diagram with top quark being
involved, as shown in Eq.~(\ref{eq:A},\ref{eq:A1},\ref{eq:A2b}).
Existence of opposite signs between the contributions of the CP-even
Higgs $H$ (negative) and the CP-odd Higgs $A$ (positive) one-loop
diagrams~\cite{Davidson:2010xv} leads to a cancelation effect,
however, it does not occur as long as $m_A>m_H$. With a large
coupling $\rho_{tt}$ and under the limit $\sin\alpha\sim 1$, the
Barr-Zee diagram involving top-quark composes the dominant
contribution (positive) to $\tau \rightarrow \mu\gamma$.

Additionally, like the case of $\delta a_\mu$, as $m_a$ is smaller
and $\theta$  is larger, the importance of the one-loop where $a$ is involved to BR$(\tau
\rightarrow \mu\gamma)$ enhances as shown in the bottom-right panel
of Fig.~\ref{Fig:results}. As well, a smaller $\alpha$ would
also induce an enhancement of the contribution of the $h$ mediated
Barr-Zee diagram with a $W$ loop.
Thus due to the contributions of this Barr-Zee diagram
and the $a$ one-loop diagram, for a region with larger $\theta$ and smaller
$\alpha$, our calculation would predict an even larger BR$(\tau
\rightarrow \mu\gamma)$ (see the top-right panel).

The theoretical prediction on BR$(\tau \rightarrow \mu\gamma)$ with
respect to $\rho_{\mu\tau}$(plotted at the top-left and middle
panels) shows that its behavior is similar to that for $\delta
a_\mu$.

It is noteworthy that since the the form factor $\mathcal{A}$ is
related to tau lepton and top quark through the concerned loops,
(see Eq.~(\ref{eq:A1},\ref{eq:A2b})), the current experimental bound
${\rm BR}(\tau\rightarrow \mu\gamma)<4.4\times
10^{-8}$~\cite{Hayasaka:2007vc,Aubert:2009ag} constrains
$\rho_{\tau\tau}$ and $\rho_{tt}$ strictly. The red region in the
bottom-left panel of Fig.~\ref{Fig:results} is allowed. It is
interesting to note that  $\rho_{tt}$ and $\rho_{\tau\tau}$ should
have opposite signs as favored by the data. This result agrees with
that given by the authors of Ref.~\cite{Omura:2015nja}. One can see
that the upper bound demands $\rho_{\tau\tau}$ to be small as about
$|\rho_{\tau\tau}|<0.04$.

\item \textit{DM relic density and GCE}

As long as assuming that the observed GCE is caused by dark matter
annihilation, the lower bound of the annihilation cross section
$\langle\sigma v_r\rangle$ should be about $0.5\times 10^{-26} $cm$^3$/s as
discussed in Ref.\cite{Calore:2014nla}. The upper bound of $\langle\sigma
v_r\rangle$ is determined to be $4.0\times 10^{-26}$ cm$^3$/s at $95\%$
CL.~\cite{2015arXiv150302632T}. Also the data of Pass 8 of Fermi-LAT~\cite{Ackermann:2015zua} from dwarf spheroidal satellite galaxies set a new upper bound
$(\sim1.0\times 10^{-26} $cm$^3$/s) on the dark matter annihilation cross section at the dark matter mass 30 GeV.
The areas in Fig.~\ref{Fig:results} that give an annihilation cross section of $\langle\sigma v_r\rangle=0.5-4.0(1.0)\times 10^{-26} $cm$^3$/s are depicted as the cyan (orange) regions.
As for the top-left (bottom-left) parameter space, the calculated values of  $\langle\sigma v_r\rangle =1.8(2.5)\times10^{-26}$cm$^{3}$/s.

The blue curves and contours from Fig.~\ref{Fig:results} represent the correct dark matter relic density ($\Omega h^2=0.1197\pm0.0022$~\cite{Ade:2015xua}). We note that the range of $\langle\sigma
v_r\rangle$  favored by the DM relic density and GCE as expected highly
depends on $m_a$, $\theta$, and $\rho_{bb}$  (see
Sec.~\ref{Darkmatter}). Due to the enhancement of the annihilation
cross section when the mediator mass  $m_a$ is close to $\sim
2m_\chi$, the dark matter relic density  rules out a range in the
parameter space (see the bottom-right panel).

With a sizeable $\rho_{bb}\sim O(0.1)$, the DM relic density  and
GCE could be tolerated simultaneously  in a range of $\langle\sigma v_r\rangle
=(0.5-4)\times 10^{-26}$cm$^{3}$/s.

However, the results newly reported by the Fermi-LAT and DES collaborations 
constrain  $\langle\sigma v_r\rangle$ to be smaller than $1.0\times 10^{-26}
$cm$^3$/s for the 30 GeV dark matter fermions, which is slightly
smaller than the value required by the thermal relic abundance, thus
there should exist other additional DM annihilation channels to
make up  the correct DM relic
density~\cite{Cheung:2014lqa,Bi:2015qva}.

\item{\textit{The CMS constraints}}
\label{cmss}

\begin{table*}[!htp]
 \begin{center}
  \begin{tabular}{c|c|c|c|c|c|c|c} \hline
  $$ & $\text{CMS}$\cite{Khachatryan:2014jba} & $m_a-\theta$ & $\alpha-\theta$ & $\alpha-\rho_{\mu\tau}$ & $\rho_{\mu\tau}-\theta$ & $\rho_{\tau\tau}-\rho_{tt}$ & $\rho_{\mu\tau}-\rho_{bb}$ \\ \hline
   $\mu_\tau$  &~$0.91\pm0.28$~ &~0.94-1.07~& ~0.92-1.11~ & ~0.93-1.10~ & ~0.94-1.08~& ~0.85-1.19~ &~0.92-1.0~ \\ \hline
   $\mu_b$     &~$0.84\pm0.44$~ &~0.58-0.67~& ~0.63-0.65~ & ~0.64-0.65~ & ~0.58-0.67~& ~0.64-0.69~ &~0.64-0.66~ \\ \hline
   $\mu_W$     &~$0.83\pm0.21$~ &~0.88-1.0~& ~0.88-1.04~ & ~0.89-1.03~ & ~0.88-1.0  ~& ~0.96-1.04~ &~0.87-1.03~ \\ \hline
   $\mu_Z$     &~$1.0\pm0.29$~ &~0.88-1.0~& ~0.88-1.04~ & ~0.89-1.03~ & ~0.88-1.0  ~& ~0.96-1.04~ &~0.87-1.03~ \\ \hline
   $\mu_\gamma$&~$1.12\pm0.24$~ &~0.88-1.0~& ~0.88-1.04~ & ~0.89-1.04~ & ~0.88-1.01 ~& ~0.97-1.04~ &~0.88-1.04~ \\ \hline
  \end{tabular}
 \end{center}
\caption{A comparison of  the signal strengths in the allowed parameter spaces of this model with the Higgs data coming from CMS~\cite{Khachatryan:2014jba}.}
\label{tab:cms}
\end{table*}

Recently, the CMS collaboration combined the comprehensive sets of production
and decay measurements for the 125 GeV-Higgs boson, including decay channels into $\gamma\gamma, ZZ^*, WW^*,
\tau^+\tau^-$, $b\bar b$, and $\mu^+\mu^-$, and found no significant deviation
from the standard model predictions~\cite{Khachatryan:2014jba}.
That synthesis should severely constrain the parameter spaces of all built models.

More specifically, the signal strengths for $ \tau^+ \tau^- , b \bar b
  , W W^\ast , Z Z^\ast , \gamma\gamma$ channels are  defined as
  \begin{equation}
  \mu_i = \frac{ \sigma (h)\times \text{BR}_i } {\sigma^{\text{SM}}(h) \times \text{BR}^{\text{SM}}_i}\; ,
  \end{equation}
which can be gained by fitting the CMS data within 1$\sigma$ tolerance~\cite{CMS:2014ega}. In the text
$ \sigma(h)$ and $\sigma^{\text{SM}}(h)$ ( $\text{BR}_i$ and BR$^{\text{SM}}_i$) correspond to the Higgs production cross section (decay branching fractions of the five decay modes) predicted by respectively this model and the SM as BR$ _i=\Gamma(h\rightarrow i i )/(\Gamma_{\text{BSM}}(h)+\Gamma_{\text{SM}}(h))$.

We find that the upper bound on $\mu_W$ and the lower bound on $\mu_\gamma$ set more rigorous constraints on the parameters of this model, and the grey areas in Fig.\ref{Fig:results} are excluded. With the allowed spaces of the parameters, the ranges of the signal strengths $\mu_i$  are given in Table.~\ref{tab:cms}.

The CMS Higgs decay signal strength strongly constrains the $h/H$, $a/A$ mixing angles $\alpha$ and $\theta$, concretely it demands both $\sin\alpha$ and  $\cos\theta$ to be close to unity (see the middle-left and top panels). The constraint on $\alpha$ is found to be compatible with
the parameter space obtained by accounting for the $h\rightarrow \mu\tau$ excess.
$\rho_{bb}$ is also severely bounded since it could
affect the primary decay mode of the SM-like Higgs significantly. As is shown in middle-right panel of Fig.~\ref{Fig:results}, which is a contour diagram of $\rho_{bb}$-$\rho_{\mu\tau}$, one notes that only a small band with
$\rho_{bb}$ being between $-0.15$ and $-0.22$ is not ruled out.

Meanwhile from the contour diagram of $m_a-\theta$, one can see that as long as BR$(h\rightarrow aa)$ is larger than $0.12$, the grey regions in the figure are excluded by the CMS Higgs signal strengths.

For the not-yet-detected channel $h\to aa$ , the pseudoscalar primarily decays into $b \bar b$, then there should be an additional contribution of the process $h\to aa\to 4b$ in the $h\to b \bar b$ searches \cite{Curtin:2013fra}, and the data of the CMS experiment \cite{Chatrchyan:2013zna,Khachatryan:2014jba} would definitely constrain the coupling between $hb\bar b$ as long as $m_a<m_h/2$. The upper bound on the undetected decay of Higgs will be further improved as the $b\bar b$ pair production is measured at the 13 (14)TeV. Moreover the $h \to 2b 2\nu$ searches at LHC \cite{Ipek:2014gua} could give more rigorous constraints on  the parameters $m_a$ and $\theta$.

\item{\textit{Collider observation of the pseudoscalar}}
\label{lhcpre}

For the benchmark scenarios of the work, when $m_a>2 m_\chi$, the masses relation $m_A>m_h+m_a$ opens the 
channel $pp(gg)\rightarrow A\rightarrow h(h\rightarrow\gamma\gamma)a(a\rightarrow \chi\bar{\chi})$ for the observation of the pseudoscalar $a$ at the LHC, which is found to be the case of the mono-Higgs searches in \cite{No:2015xqa}. 
The backgrounds are dominated by the SM process $pp\rightarrow Z\gamma\gamma$\footnote{Here, we would like to mention that including
the background $Z\gamma+jets$ (with a jet faked a photon) might reduce the observation significance a bit} with $Z\rightarrow\nu\nu$, 
and the Higgs associated production process $pp\rightarrow Zh$ with $Z\rightarrow\nu\nu$.
The collider analysis is performed by generating signal and background events with {\sc MadGraph5$\_$aMC$@$NLO} \cite{Alwall:2014hca,Alwall:2011uj}  at 14 TeV, 
 and then passing on to {\sc Pythia~8.1}\cite{Sjostrand:2007gs} for parton shower and hadronization, and the detector simulation is conducted by {\sc Delphes~3}\cite{deFavereau:2013fsa} at last. After implement the selection cuts  $m_{\gamma\gamma}\in [120, 130]$ GeV, as well as $E\hspace{-0.08in}\slash_T,P_T^{\gamma\gamma}>76$ GeV following Ref.\cite{No:2015xqa}, the number of events of signal and background are obtained as $37$ and $48$ respectively, thus the significance for the observation of $a$ is found to $S/\sqrt{S+B}\sim 4$ for the benchmark of $m_a-\theta$ with $\theta=0.08, m_a=76$ GeV and with an integrated luminosity $\mathcal{L}=300 fb^{-1}$.
 
The probe of the pseudoscalar could also be conducted with hard b-jets and large missing transverse energy~\cite{Lin:2013sca} when $m_a>2m_\chi$ and the pseudoscalar dominantly decay to dark matter. And when the pseudoscalar decay mostly to b-quarks, the collider search of the pseudoscalar could be found in Ref.\cite{Fan:2015sza}.

\end{enumerate}

\section{Conclusions and Discussions}

The discovered anomalies by the LHC experiments and observations of the Dark matter greatly excite the curiosity of human beings and inspire enthusiasm of searching
for new physics beyond standard model. However, so far, the trend is not very successful. Even though we know new physics must be around, but do not know its scale.
Direct and indirect search for dark matter, LHC experiments, long-baseline and short-baseline neutrino experiments and numerous lower energy experiments including BELLE, BES
and many others provide hints towards new physics beyond standard model, however at the same time, set more and more rigorous constraints on those models which have been proposed
to explain the anomalies observed in astronomy and high energy experiments. Some of the models survive the so-far measurements and many have been ruled out.

In this work, we have extended the Type-III 2HDM by introducing a pseudoscalar $a$ which can mediate the DM pair annihilation process.
In a recent paper, Han et al.\cite{Han:2015yys} also extended the 2HDM with an aligned Yukawa sector to explain the $(g-2)_{\mu}$ excess, in comparison, our scheme is somewhat different from theirs. We not only
consider the $(g-2)_{\mu}$ excess, but also many other constraints from both earth experiments and cosmology.
In this framework, the LFV process $h\rightarrow\mu\tau$ observed at LHC is addressed, the possibility of explaining the muon $g-2$ anomaly, dark matter relic abundance and GCE
have also been investigated, the role played by the newly introduced $a$ 
is studied in some details.
It is found that there indeed exist certain parameter spaces which can
tolerate $h\rightarrow \mu\tau$ excess, muon $g-2$ discrepancy and dark matter relic abundance.

With the flavor violating coupling $\rho_{\mu\tau}\sim O(0.1)$ and a tiny mixing between $h$ and $H$ around $\sin\alpha\sim1$, the observation of the $h\to\mu\tau$ excess could be easily accommodated. The pseudoscalar $a$ opens an important undetected decay channel for the SM-like Higgs ($h\to aa$) as $m_a<{m_h/ 2}$, thus affects the branching ratio of $h\to \tau\mu$. It
also plays a role to explain the discrepancy between theoretical prediction and data of $(g-2)_{\mu}$.
A smaller $m_a$ and a slightly larger CP-odd $a/A$ mixing $\theta$ helps to interpret the muon $g-2$ anomaly. Increasing the CP-odd Higgs mixing angle and $|\rho_{\mu\tau}|$
increases BR$(\tau\rightarrow\mu\gamma)$, thus would further constrain the parameters space of our model. The measurement of the branching ratio of $\tau\rightarrow\mu\gamma$
sets a stringent bound on the flavor conserving Yukawa couplings $\rho_{tt}$ and $\rho_{\tau\tau}$, moreover, it determines an opposite sign between the two couplings.
There are parameter regions in our model allowed by the current experimental data of $\tau\to\mu\gamma$ where both the measured $h\to\mu\tau$ excess and the muon g-2 anomaly can be explained.

To account for the dark matter relic density, the $a/A$ mixing angle should be of order $O(0.01)$ and $\rho_{bb}\sim O(0.1)$ is required, however this  parameter
region does not coincide with that favored by the GCE observation if it is postulated that the GCE is fully coming from the $\chi\bar\chi$ DM pair annihilation. This inconsistency may imply that  there are other sources to result in the GCE besides the pure $\chi\bar\chi$ DM annihilation mechanism.

The LFV process $h\to\mu\tau$ is studied in terms of our model and our prediction on  BR$(h\to\mu\tau)$ is qualitatively consistent with that
observed by CMS and ATLAS at 8 TeV, however to make a decisive conclusion, more data are needed and
the LHC Run II of 14 TeV should help.
A synthesis  of data accumulated by high energy collider LHC, the future SPPC of 100 TeV  and maybe some lower energy experiments as well as the new astronomical
observation  would make the whole picture clearer,
then we will be able to judge whether this model indeed works or needs to be further modified.
For the case of $m_a>2m_\chi$, the mono-Higgs search provide one possible probe of the pseudoscalar.

\appendix*

\section{Loop functions }
\label{ope}

The function $f_\phi$ and $h$ using for the calculation of
$g-2$ and $\tau\rightarrow \mu\gamma$ are given by,
\begin{eqnarray}
f_{A,a}(r)=\frac{r}{2}\int_0^1 dx\frac{1}{x(1-x)-r}\log\frac{x(1-x)}{r}\;  .
\label{opef}
\end{eqnarray}

The functions $f\left(x,y,r\right)$ and Inami-Lim function $Y(x)$ being used in
Eq.~(\ref{Bsmumu}) are
\begin{align}
f\left(x,y,r\right)&=\frac{x}{8}\Bigg[-\frac{r\left(x-1\right)-x}{\left(r-1\right)\left(x-1\right)}\log r+\frac{x \log x}{\left(x-1\right)}
\nonumber
\\
&\quad-\frac{y \log y}{\left(y-1\right)}+\frac{x \log y}{\left(r-x\right)\left(x-1\right)}\Bigg],\\
Y\left(x\right)&=\frac{x}{8}\left[\frac{x-4}{x-1}\log x +\frac{3x \log x}{\left(x-1\right)^2}\right].
\label{bsf}
\end{align}

\begin{acknowledgments}
We are grateful to Peng-Fei Yin and Zuowei Liu for helpful discussions on Galactic Center gamma ray excess.
This work is supported by the National Natural Science Foundation of China under Grants No. 11375128, and No. 11547305,
and by Fundamental Research Funds for the Central Universities under Grant No. 0903005203404.
\end{acknowledgments}

\end{document}